\documentclass{PoS}

\usepackage{wrapfig}
\def\GeV{{\rm GeV}}
\def
\TeV{{\rm TeV}}

\title{Inclusion of new LHC data in MMHT PDFs}

\ShortTitle{MMHT}

\author{\speaker{R.S.~Thorne} \\
        Department of Physics and Astronomy, \\
        University College London, WC1E 6BT, UK\\
        E-mail: \email{robert.thorne@ucl.ac.uk}}

\author{L.A.~Harland-Lang\\
        Department of Physics and Astronomy, \\  
        University College London, WC1E 6BT, UK\\
        E-mail: \email{l.harland-lang@ucl.ac.uk}}

\author{A.D.~Martin\\
        Institute for Particle Physics Phenomenology,\\
        University of Durham, DH1 3LE, UK \\ 
        E-mail: \email{A.D.Martin@durham.ac.uk}}

\abstract{I consider the effects of including a variety of new LHC data sets 
into the MMHT approach for PDF determination. 
I consider the impact of fitting new LHC 
and Tevatron data, which leads to clear improvements in some PDF uncertainties. 
There are specific issues with ATLAS 7 TeV jet data and I  include a  
discussion of the treatment of correlated uncertainties and briefly the 
effects of NNLO corrections. I also present preliminary results with the 
inclusion of the high precison final ATLAS 7 TeV $W,Z$ rapidity-dependent data.}

\FullConference{XXV International Workshop on Deep-Inelastic Scattering and Related Subjects\\
		3-7 April 2017\\
		University of Birmingham, UK}

\begin{document}

The MMHT2014 parton distributions \cite{Harland-Lang:2014zoa}
were the first from our group to include LHC data in their determination. 
Soon after we considered the effect of updating these to
include the final HERA total cross section measurements 
\cite{Abramowicz:2015mha}, noting only minor changes in the central values of these PDFs and reductions
in uncertainties of up to $10\%$ \cite{Harland-Lang:2016yfn}. 
I will start from the baseline of the PDFs in \cite{Harland-Lang:2016yfn}
when considering the effect of further updates in this account.  
 
The MMHT PDF fit has been updated to account for a fit to a wide variety of new hadron collider data. We include in the PDF determination 
high rapidity $W,Z$ data from LHCb at $7$ and 
$8~\TeV$ \cite{Aaij:2015gna,Aaij:2015zlq,Aaij:2015vua}, 
$W+c$ jets from CMS \cite{Chatrchyan:2013uja}, which constrains 
strange quarks, high precision CMS data on $W^{+,-}$ rapidity 
distributions \cite{Khachatryan:2016pev} which can also be interpreted as an asymmetry measurement, 
and also the final $e$ asymmetry data from D0 \cite{D0:2014kma}. 
All these cross sections are calculated at NLO using MCFM~\cite{Campbell:2015qma} in combination with Applgrid~\cite{Carli:2010rw} and FEWZ~\cite{Li:2012wna}. 

\vspace{-0.2cm}

\begin{table}[h]
\begin{center}
\begin{tabular}{|c|c|c|c|}
\hline
 &Points& NLO $\chi^2$&NNLO $\chi^2$\\ \hline
$\sigma_{t\overline{t}}$ &18&19.6 (20.5)&14.7 (15.3)\\ 
LHCb 7 TeV $W+Z$&33 &50.1 (45.4)&46.5 (42.9)\\
LHCb 8 TeV $W+Z$&34&77.0 (58.9)&62.6 (59.0)\\
LHCb 8 TeV $Z\to ee$ &17&37.4 (33.4)&30.3 (28.9)\\
CMS 8 TeV $W$&22&32.6 (18.6)&34.9 (20.5)\\
CMS 7 TeV $W+c$&10&8.5 (10.0)&8.7 (7.8)\\
D0 $e$ asymmetry&13&22.2 (21.5)&27.3 (25.8)\\
\hline
Total&3405 (3738)&4375.9 (4336.1)&3741.5 (3723.7)\\
\hline
\end{tabular}
\vspace{-0.1cm}
\caption{$\chi^2$ at NLO and NNLO for the prediction (fit) to the new LHC and Tevatron data included in the MMHT -- 2016 fit. Also shown is the total number of points without (with) the new data included.}
\label{tab:Tab1}
\end{center}
\end{table}

\vspace{-0.7cm}

The results are shown in Table 1. The predictions from the existing PDFs are generally good, and there is no real tension with other data when refitting (at NLO $\Delta \chi^2 =9$ for the remainder of the data
and at NNLO $\Delta \chi^2 =8$). The fit quality is slightly ($\sim 10$ units) better than in a previous report \cite{Harland-Lang:2016zfc}
due to improvements (and one correction) in NNLO K-factors. 
The data which requires most PDF adjustment is the new 8~TeV CMS $W^{\pm}$ rapidity
and asymmetry data. This is shown in the left of Figure 1 
where good agreement is seen after refitting. (We fit to individual distributions not the asymmetry, 
but it is easier to display the latter.)
We have also included further results on 
$\sigma_{\bar t t}$ to those in the MMHT2014 study. The NNLO K-factors are calculated using \texttt{top++}~\cite{Czakon:2011xx}. The fit quality is very good and with $\alpha_S(M_Z^2)=0.118$ the fitted 
$m_t^{pole}=173.4~\GeV$ at NNLO, and at NLO 
$m_t^{pole}=170.2~\GeV$. In contrast the MMHT values were $m_t^{pole}=174.2~\GeV$ and $m_t^{pole}=171.7~\GeV$. When the coupling is left free in our new fits then at NLO $\alpha_S(M_Z^2)$ 
stays very close to the MMHT2014 value of 0.120 but at NNLO $\alpha_S(M_Z^2)$ is
marginally above 0.118, slightly higher than the value of 0.1172 in MMHT2014 \cite{Harland-Lang:2015nxa}.

\begin{figure}[]
\vspace{-0.0cm}
\centerline{\includegraphics[width=0.44\textwidth]{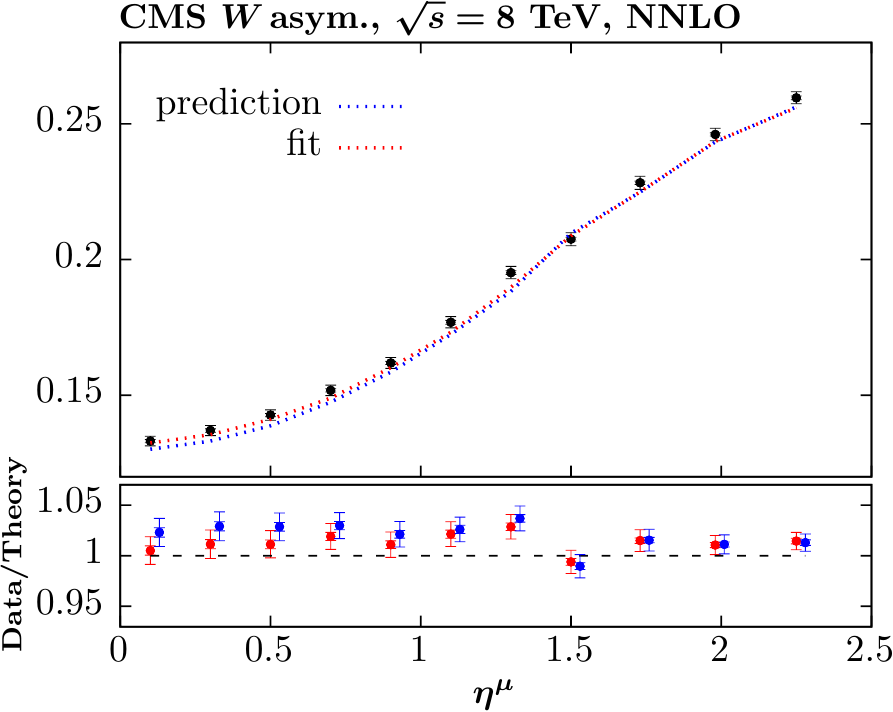}
\includegraphics[width=0.56\textwidth]{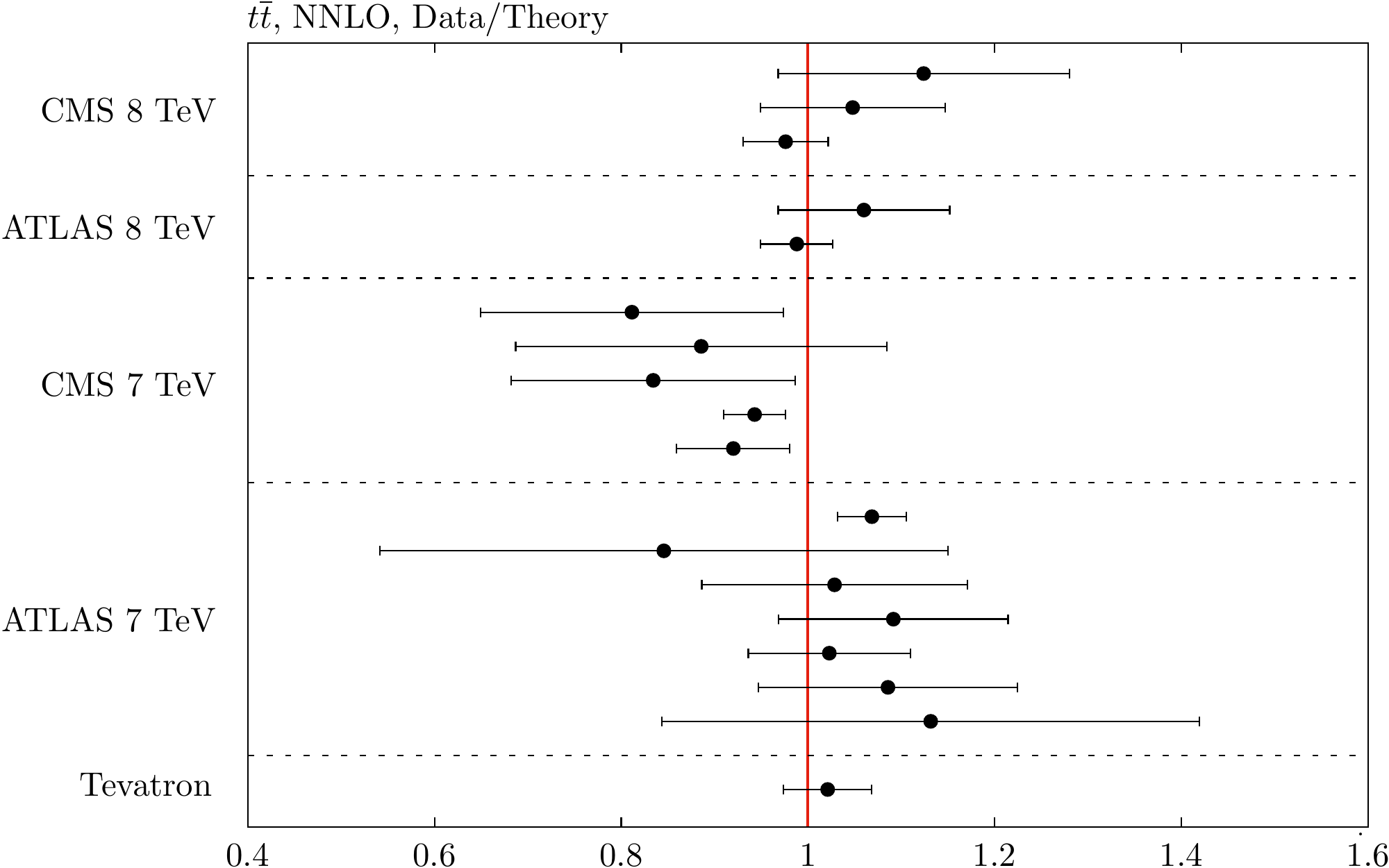}}
\vspace{-0.2cm}
\caption{Good agreement with new 8~TeV CMS $W^{\pm}$ rapidity
and asymmetry (left).
The data/theory ratio for more up-to-date results on $\sigma_{\bar t t}$ (right)}
\vspace{-0.6cm}
\label{Fig1} 
\end{figure}

We have generated a central set at 
NLO and NNLO for the fit including these new data -- labelled MMHT (2016 fit).
We also generate PDF eigenvector sets for uncertainties at NNLO
using the same basis of free PDF parameters as in MMHT2014. Hence, there are 
50 eigenvector directions, and 
14 of these are best constrained by one of the new (LHC)
data sets. There is a large reduction in the $s+\bar s$ uncertainty, but little change 
in the central value,  due to the $W+c$ jets data. There is also a significant change 
in $u_v- d_v$ at small-$x$ from the CMS $W$ data,  
and noticeably reduced uncertainty with the new data inclusion.

We have also attempted a NLO fit including the high luminosity ATLAS $7~\TeV$ inclusive jet data 
\cite{Aad:2014vwa}. Full details have already been presented in \cite{Harland-Lang:2017dzr}, so we 
present a brief summary here. The MMHT prediction gives $\chi^2/N_{pts}=413.1/140$, and a refit gives 
an improvement only to $\chi^2/N_{pts}=400.4/140$.  We cannot simultaneously fit 
data in all the rapidity bins. 
The data set has large correlated systematics which dominate over uncorrelated 
uncertainties. 
The best possible fit requires a shift in data against theory which is different 
from one rapidity bin to another, and hence not allowed due to the correlations of the 
uncertainties between bins, but a good fit ($\chi^2/N_{pts} \sim 1$) is possible when fitting 
each individual rapidity bin separately. 
Hence, we look at the shifts due to each source of correlated uncertainty.  
A small number of sources prefer very different values when 
fits to different bins are performed. Hence, we consider 
fits to all data when decorrelating some error sources, i.e. making them 
independent between the 6 rapidity bins.  There is a very significant improvement, 
particularly from decorrelating source jes21.  
With correlations between rapidity bins relaxed for just two 
sources of systematics, jes21 and jes62,  $\chi^2/N_{pts}=178/140=1.27$. 

\begin{figure}[]
\vspace{-0.2cm}
\centerline{\includegraphics[width=0.48\textwidth]{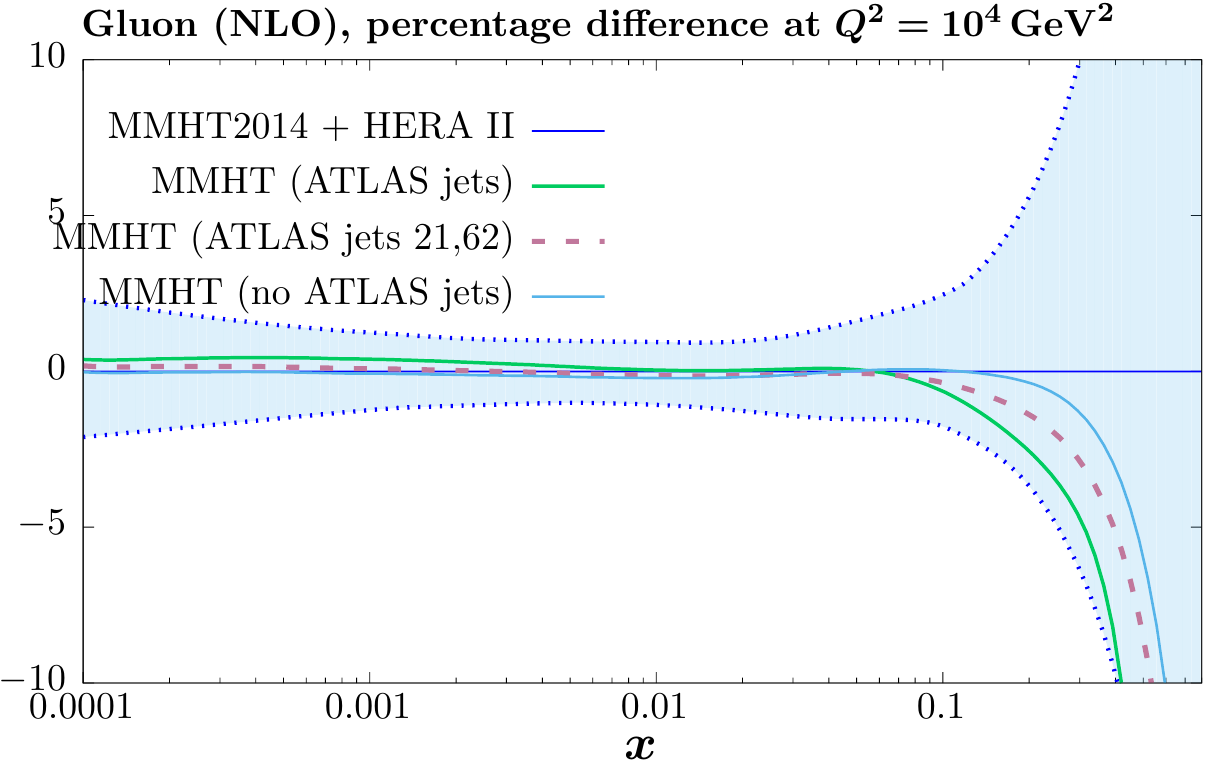}
\includegraphics[width=0.48\textwidth]{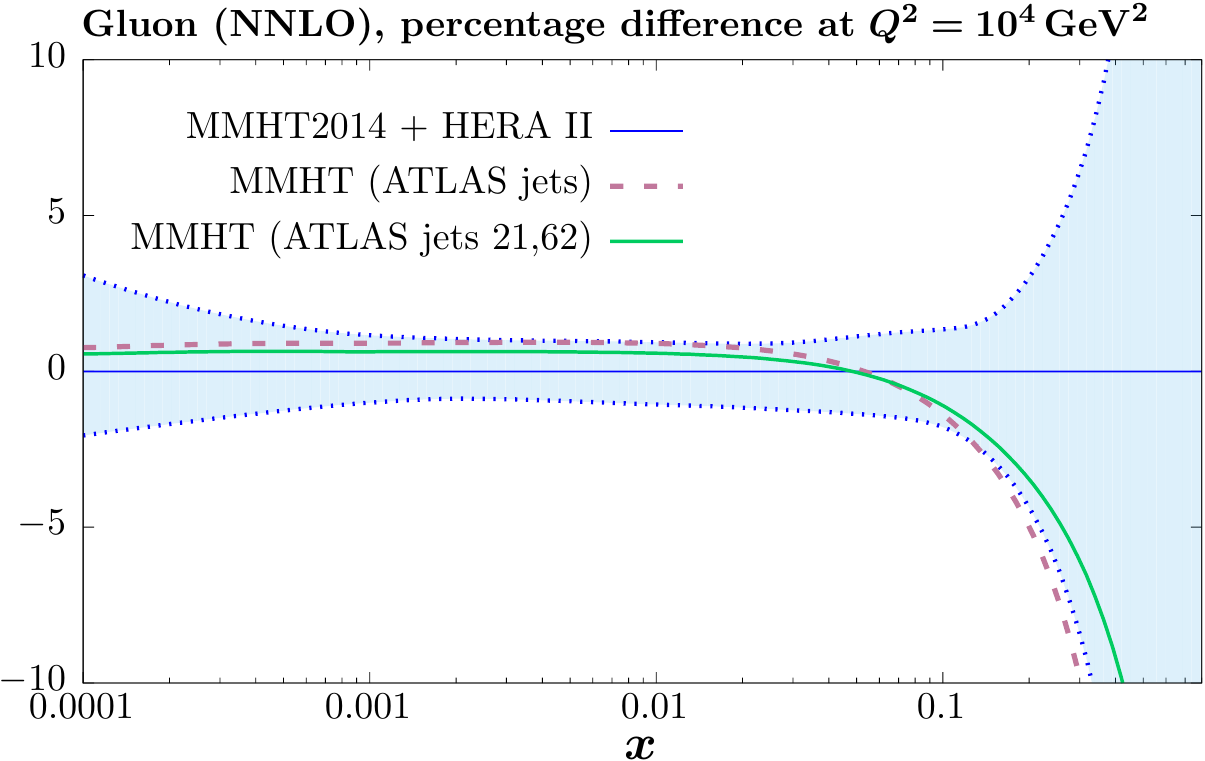}}
\vspace{-0.4cm}
\caption{The gluon at NLO (left) and NNLO (right) when ATLAS jet data is fit.}
\vspace{-0.4cm}
\label{Fig2} 
\end{figure}

We also consider the effect of NNLO corrections to jet cross sections \cite{Currie:2016bfm}. 
We find a significant, if not dramatic, deterioration in the fit quality, which might be expected as the 
NNLO corrections move the unshifted data and theory slightly further apart.  
The gluon obtained when including ATLAS jet data at NLO and NNLO is shown in Figure 2.
The effect on the best fit gluon is noticeable, but within 
(or at the boundary) of the uncertainties. It is softer at very high $x$ and there is a slightly 
smaller effect at NLO than at NNLO. 
The result on the gluon is not very dependent on whether uncertainty sources are decorrelated 
or not even though the fit quality is vastly different.

\begin{figure}[]
\vspace{-0.2cm}
\centerline{\includegraphics[width=0.48\textwidth]{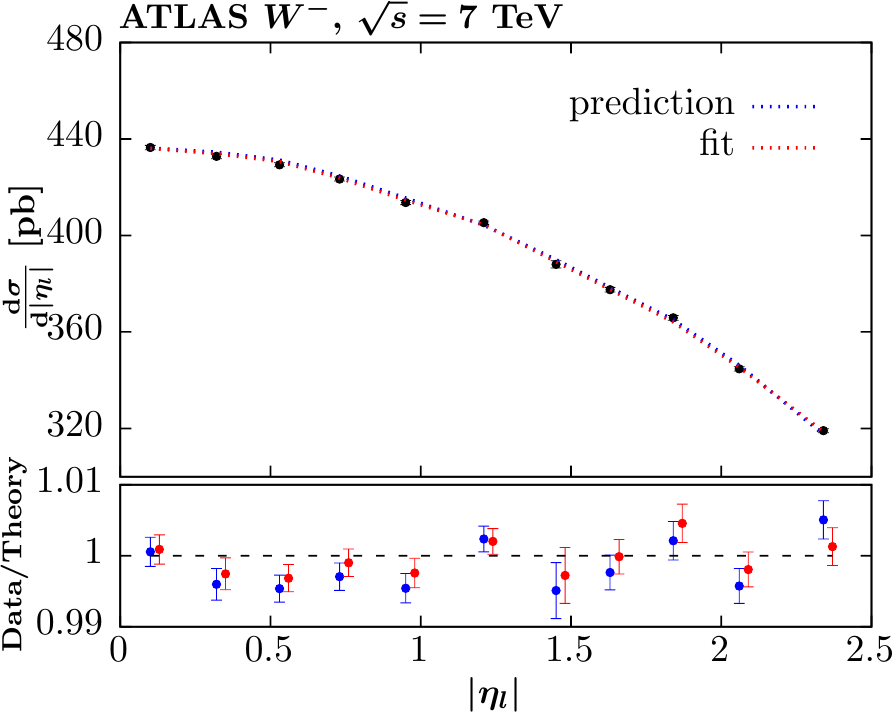}
\includegraphics[width=0.48\textwidth]{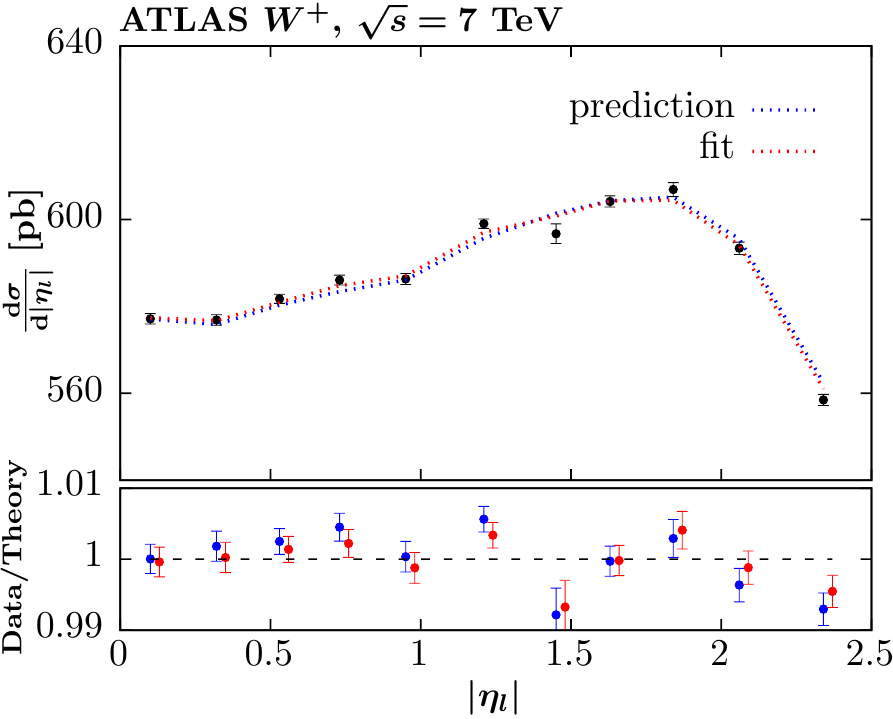}}
\centerline{\includegraphics[width=0.48\textwidth]{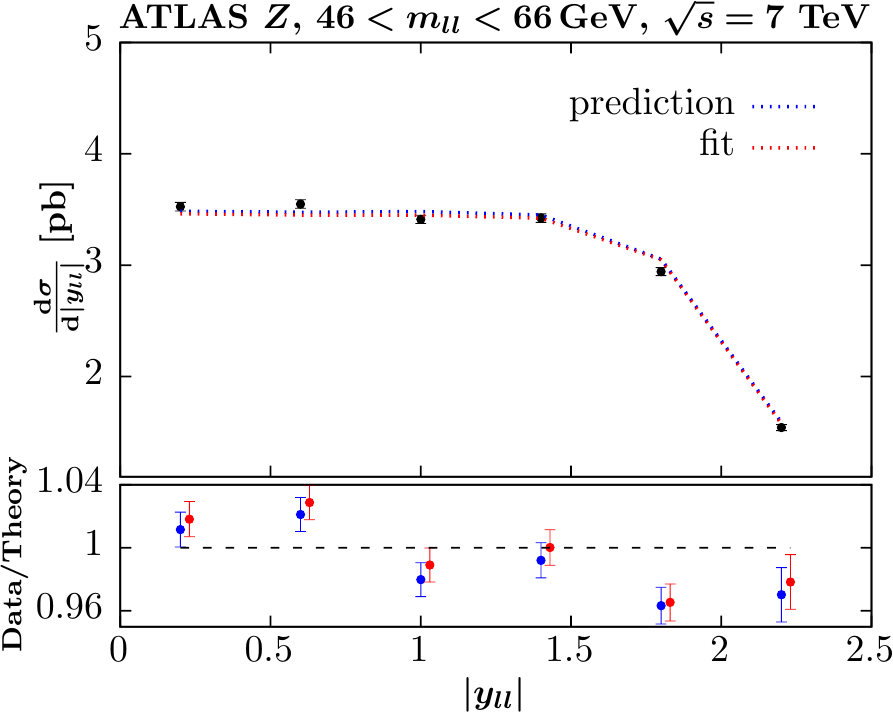}
\includegraphics[width=0.48\textwidth]{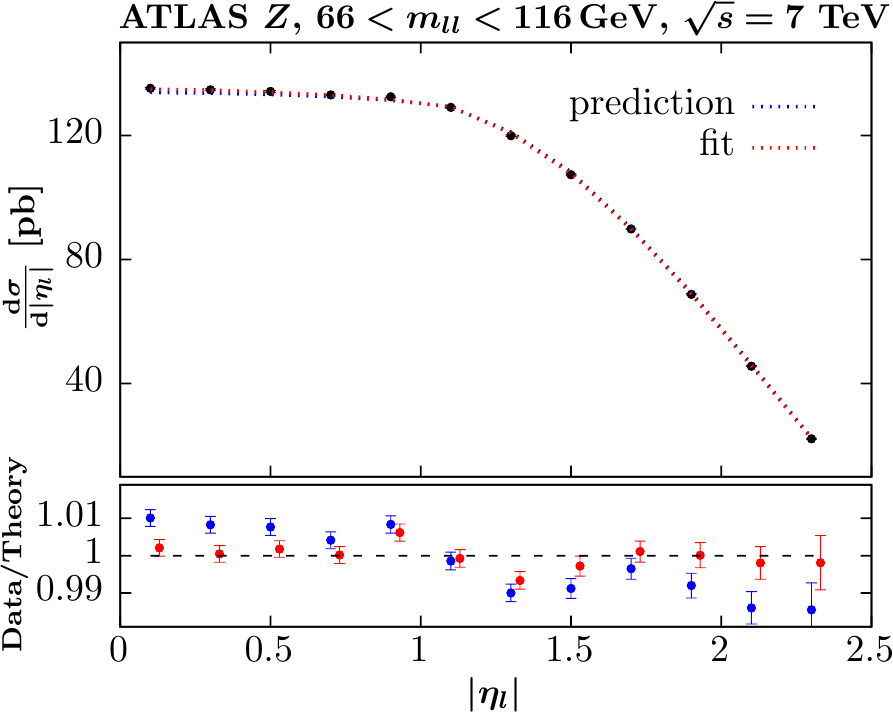}}
\vspace{-0.2cm}
\caption{Fit to ATLAS $W,Z$ data. A slight reduction in theory for $W^-$ 
is required (top left) and the opposite 
for $W^+$ (top right). There is a significant change in the shape required for $Z$ production (bottom right), i.e. the 
theory becomes 
higher at low $\vert \eta \vert$ and lower at high $\vert \eta \vert$.}
\vspace{-0.8cm}
\label{Fig3} 
\end{figure}

Finally, we consider the inclusion of recent high precision ATLAS $W,Z$ data at 7~TeV \cite{Aaboud:2016btc}.
We obtain $\chi^2/N_{pts}\sim 400/61$ from MMHT14 PDFs at NNLO (though the $\chi^2$ lowers significantly when PDF errors are included 
\cite{Aaboud:2016btc}).
For PDFs with final HERA combined data (and 
some new $\sigma_{t\bar t}$ data points) we obtain $\chi^2/N_{pts}\sim 387/61$. 
We use this as our ``baseline''.  
Including the ATLAS $W,Z$ data in our fit the quality reduces to $\chi^2/N_{pts}\sim 130/61$, similar to the result for MMHT PDFs in \cite{Aaboud:2016btc} when PDF profiling is used. The comparison of theory to data is shown in Figure 3.
The deterioration in the fit to other data is $\Delta \chi^2=54$. This is worst for CMS double 
differential $Z/\gamma$ data ($\Delta \chi^2=17$) and
CCFR/NuTeV dimuon data ($\Delta \chi^2=16$). For the latter the charm meson to muon branching 
ratio requires a $25\%$ shift, but this is not exceptional as the uncertainty is $15\%$. 
Other data sets showing deterioration are fixed target DIS data, E866 Drell-Yan asymmetry and CDF $W$-asymmetry data. We also increase the weight of new ATLAS data by a factor of 10, and 
the $\chi^2$ improves to $\chi^2/N_{pts}\sim 121/61$. 
The deterioration in the fit to other data becomes $\Delta \chi^2=92$. There is a further increase in CMS double differential $Z/\gamma$ data ($\Delta \chi^2=24$) and  E866 Drell-Yan asymmetry (the dimuon data is not any worse). 
There is now also a deterioration in HERA combined data and CDF differential $Z/\gamma$ data. 
We also perform a fit with scales set to $\mu_{R,F} =M_{W,Z}/2$
rather than $\mu_{R,F} =M_{W,Z}$. 
As in \cite{Aaboud:2016btc} we find a reduction in $\chi^2$ to $\chi^2/N_{pts}\sim 106/61$, spread over data subsets. There is most noticeable improvement for $W^+$ data,
with some small improvement for the lower mass $Z/\gamma$ data and 
less fluctuation for the $Z$ peak rapidity distribution.

We also perform a fully updated fit with all the new LHC data
mentioned (except the jet data).  The simultaneous inclusion of the ATLAS 
$W,Z$ data lowers the $\chi^2$ for the other
new LHC (plus final D0) data by $\Delta \chi^2=-10$, while the other data 
in the fit sees 
little change, i.e. $\Delta \chi^2=3$ in total, with essentially no change in 
ATLAS $W,Z$ data.
Hence, the ATLAS $W,Z$ data and other new LHC data are fully compatible and 
any pulls tend to be in the same direction. Only the CMS $W+c$ fit deteriorates very slightly. We  
generate PDF eigenvector sets for uncertainties at NNLO
using the same basis as in MMHT2014. Of the 50 eigenvector directions, 
21 are best constrained by one of the new LHC
data sets. 
There is a large increase in $s+\bar s$ and a decrease in its uncertainty.
The correlation with the fit to dimuon data (i.e. lower branching ratio) leads to a necessary 
increase in the cross section at all $x$. 
For $x>0.1$ this process has a significant down quark contribution despite 
Cabibbo suppression since $d(x>0.1,Q^2) \gg s(x>0.1,Q^2)$. Since the down 
quark is well constrained, the enhanced cross section is obtained by a very 
large increase in strange quark for $x\sim 0.1$. The large change in the 
charm meson to muon branching ratio may, however, be mitigated by 
NNLO corrections to dimuon 
production, which appear to be negative, particularly at smaller $x$ \cite{Berger:2016inr}. Implementing these corrections in a PDF fit will be an important development.  

\begin{figure}[]
\vspace{-0.2cm}
\centerline{\includegraphics[width=0.48\textwidth]{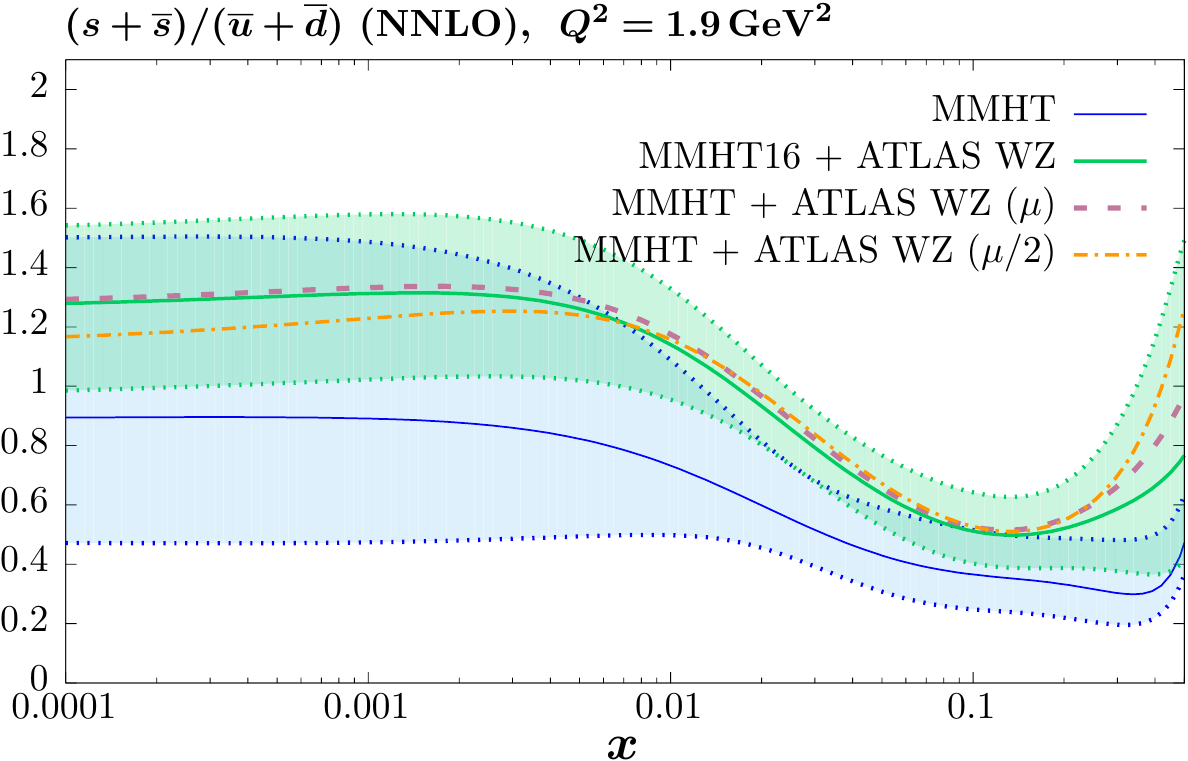}
\includegraphics[width=0.48\textwidth]{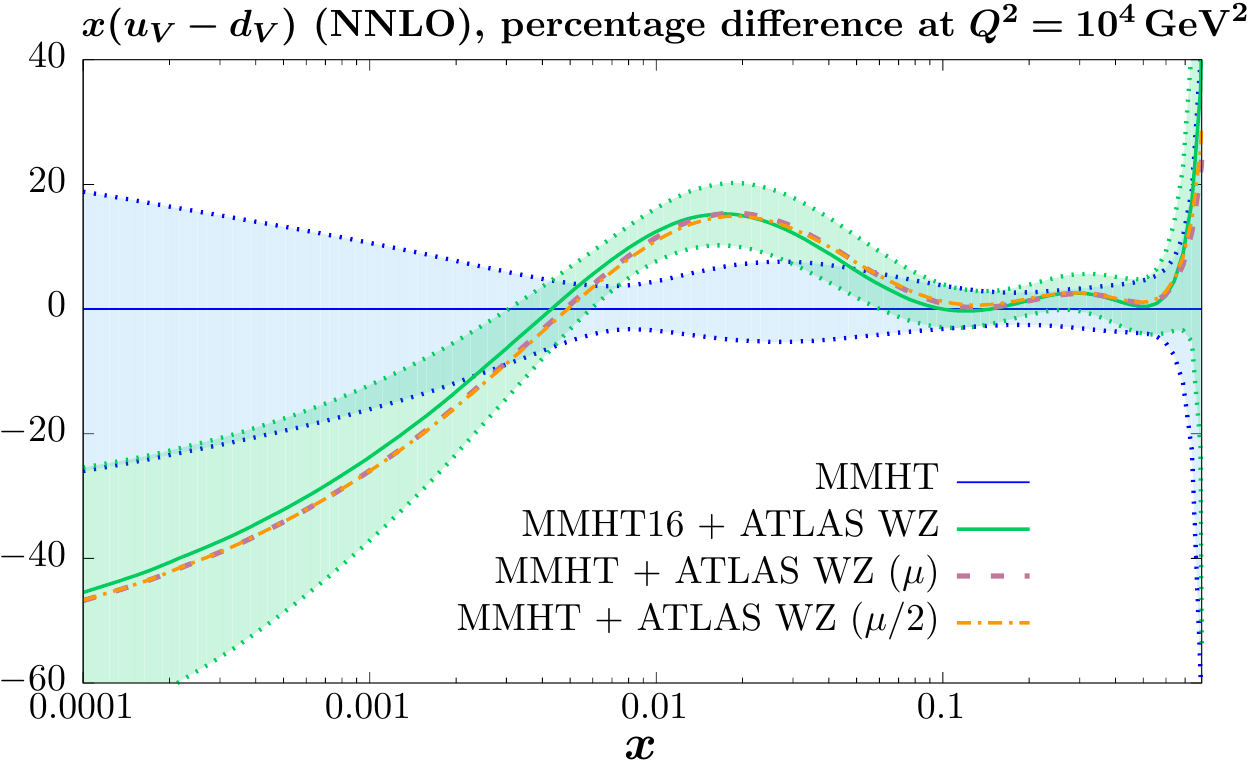}}
\vspace{-0.1cm}
\caption{The comparison of NNLO MMHT PDFs including the new ATLAS $W,Z$ data and other new LHC data 
to the existing PDFs for the strange to light sea ratio (left) and for $u_V-d_V$ 
(right).}
\vspace{-0.3cm}
\label{Fig4} 
\end{figure}

The ratio of $(s+\bar s)$ to $\bar u + \bar d$, i.e. $R_s$ at $Q^2=1.9~\GeV^2$ is shown in Figure 4. 
At $x=0.023$ $R_s \sim 0.83 \pm 0.15$, 
compared to the ATLAS result \cite{Aaboud:2016btc}
of $R_s= 1.13^{+0.08}_{-0.13}$. Conversely, we are a little larger than the NNPDF result in \cite{Ball:2017nwa}.
Our value of $R_s$ exceeds unity at lower $x$, but this is essentially an 
extrapolation and it is very consistent with a value of 1.  
Our final fit also shows a significant impact on the shape of the valence quarks. The ATLAS $W,Z$ data pulls in the same direction as the
other new LHC data. The significant change 
in $u_V-d_V$ is also shown in Fig. 4. 
The change in the strange quark affects the entire sea, making it generally larger, but the new fit shows rather little impact on the gluon distribution.

\vspace{-0.4cm}

\section*{Acknowledgements}

\vspace{-0.3cm}

\noindent We thank the
Science and Technology Facilities Council (STFC) for support via grant
award ST/L000377/1. We thank A.M. Cooper-Sarkar and V. Radescu for supplying $K$-factors for the NNLO ATLAS $W,Z$ 
cross sections.

\end{document}